\documentclass[twocolumn,pra,aps]{revtex4}
\usepackage{graphicx}

\def\duzomniejsze{<\kern-.7mm<}
\def\duzowieksze{>\kern-.7mm>}

\def\textbf#1{{\bf #1}}
\def\beq{\begin{equation}}
\def\eeq{\end{equation}}
\def\be{\begin{equation}}
\def\ee{\end{equation}}
\def\ben{\begin{eqnarray}}
\def\een{\end{eqnarray}}
\def\beqa{\begin{eqnarray}}
\def\eeqa{\end{eqnarray}}
\def\eea{\end{array}}
\def\bea{\begin{array}}
\newcommand{\bei}{\begin{itemize}}
\newcommand{\eei}{\end{itemize}}
\newcommand{\bee}{\begin{enumerate}}
\newcommand{\eee}{\end{enumerate}}

\def\>{\rangle}
\def\<{\langle}

\begin{document}

\title{Squeezing induced transition of long-time decay rate}

\begin{abstract}
{\normalsize We investigate the nonclassicality of several kinds of
nonclassical optical fields such as the pure or mixed single
photon-added coherent states and the cat states in the photon-loss
or the dephasing channels by exploring the entanglement potential as
the measure. It is shown that the long-time decay of entanglement
potentials of these states in photon loss channel is dependent of
their initial quadrature squeezing properties. In the case of
photon-loss, transition of long-time decay rate emerges at the
boundary between the squeezing and non-squeezing initial
non-gaussian states if log-negativity is adopted as the measure of
entanglement potential. However, the transition behavior disappears
if the concurrence is adopted as the measure of entanglement
potential. For the case of the dephasing, distinct decay behaviors
of the nonclassicality are also revealed.

PACS numbers: 03.67.-a, 03.65.Ud}

\end{abstract}

\author{Shang-Bin Li}\email{stephenli74@yahoo.com.cn},

\affiliation{Shanghai research center of Amertron-global,
Zhangjiang High-Tech Park, \\
299 Lane, Bisheng Road, No. 3, Suite 202, Shanghai, 201204, P.R.
China}

\maketitle

\section * {I. INTRODUCTION}

Recently, the relations between non-classicality of optical fields
and the entanglement have been intensively investigated. It is shown
that nonclassicality is a necessary condition for generating
inseparable states via the beam splitter \cite{Kim2002}. Based on
them, a measure called the entanglement potential for quantifying
the non-classicality of the single-mode optical field has been
proposed \cite{Asboth2005}. The entanglement potential is defined as
the entanglement achieved by 50:50 beam splitter characterized by
the unitary operation
$\hat{U}_{BS}=e^{\frac{i\pi}{4}(a^{\dagger}b+ab^{\dagger})}$ acting
on the target optical mode $a$ and the vacuum mode $b$.  Suppose a
quantum field is in the state $\rho_a$, then the two-mode
entanglement in the state
$\hat{U}_{BS}(\rho_a\otimes|0_b\rangle\langle0_b|)\hat{U}^{\dagger}_{BS}$
is regarded as its value of nonclassicality. Throughout this paper,
log-negativity is explored as the measure of entanglement potential
if without additional notification. The log-negativity of a density
matrix $\rho$ is defined by \cite{vidalwerner} \be
N(\rho)=\log_2\|\rho^{\Gamma}\|, \ee where $\rho^{\Gamma}$ is the
partial transpose of $\rho$ and $\|\rho^{\Gamma}\|$ denotes the
trace norm of $\rho^{\Gamma}$, which is the sum of the singular
values of $\rho^{\Gamma}$. For any single mode pure states of quantum optical
fields, their values of nonclassicality are exactly equivalent to
their mixedness at characteristic time $\gamma{t}=\ln2$ of those
quantum optical field in the vacuum environment (see
Appendix A).

When the nonclassical optical fields propagate in the medium, they
inevitably interact with their surrounding environment, which causes
the dissipation or dephasing. Therefore, how a nonclassical optical
field losses its non-classicality in the dephasing or dissipative
channel is very desirable. The photon loss channel is the simplest
one of the general gaussian channels. Any quantum states in photon
loss channel eventually asymptotically decay into the vacuum state,
which is "classical" in the field of quantum optics.

The photon-added coherent state was introduced by Agarwal and Tara
\cite{Agarwal1991}. Recently, the single photon-added coherent state
(SPACS) is experimentally prepared by Zavatta et al. and its
nonclassical properties are detected by homodyne tomography
technology \cite{Zavatta2004}. Such a state represents the
intermediate non-Gaussian state between quantum Fock state (with
zero photon-number fluctuation but random phase) and classical
coherent state (with well-defined amplitude and phase). In
Refs.\cite{Li2007,Li20071,Li20081}, we have investigated the
entanglement potential and negativity of the Wigner function of
SPACS in photon loss channel. In short time, both of them exhibit
similar behaviors. However, the Wigner functions become positive
when the decay time exceeds a threshold value. The threshold decay
time is the same for arbitrary pure or mixed nonclassical optical
fields with zero population in vacuum state \cite{Li20081}. Thus one
can not utilize the negativity of the Wigner function to quantify
the long time behavior of the nonclassicality.

One of the characteristics of the photon-loss channel is any
nonclassical states can not encounter finite-time sudden death of
nonclassicality in photon loss channel. In this paper, we
investigate the nonclassicality of several kinds of nonclassical
optical fields such as the pure or mixed single photon-added
coherent states and the superposition of coherent states in
photon-loss or dephasing channels by exploring the entanglement
potential as the measure. Different long-time decay rates of
entanglement potential are found. It is shown that the photon-loss
channel can cause three different kinds of long-time exponential
decay rates: $e^{-\gamma{t}}$ for initial quadrature squeezed
states; $e^{-2\gamma{t}}$ for initial nonclassical states without
quadrature squeezed; and $e^{-3\gamma{t}/2}$ for the boundary states
between squeezing and no squeezing states.

This paper is organized as follows: In Sec.II, the nonclassicalities
measured by entanglement potential of some kinds of nonclassical
states in photon-loss channel, the simplest gaussian channel, are
analyzed. A initial squeezing induced transition-like behavior of
the long time decay rate of the entanglement potential is revealed.
In Sec.III, The influence of the dephasing on the nonclassicality of
single-photon-added coherent states is investigated and shown the
initial SPACS can not completely loss its nonclassicality even in
the asymptotical sense. In Sec.IV, several conclusive remarks are
given. In appendix A, it is shown that the entanglement
potentials of one-mode pure states are completely equivalent to
their mixedness achieved at the threshold decay time
$\gamma{t}=\ln2$ in the photon-loss channel.

\section * {II. SQUEEZING INDUCED TRANSITION OF LONG-TIME DECAY RATE OF NONCLASSICALITY IN PHOTON LOSS CHANNEL}

Let us first briefly recall the definition of photon-added coherent
states. We are interested in the dynamical behavior of the
nonclassicality of the photon-added coherent states (PACS) in the
dissipative channel or dephasing channel. The photon-added coherent
state was firstly proposed by Agarwal and Tara \cite{Agarwal1991},
and was experimentally prepared by Zavatta et al.
\cite{Zavatta2004}. The PACS is defined by
$|\Psi_{\alpha,m}\rangle=\frac{1}{\sqrt{N(\alpha,m)}}a^{\dagger{m}}|\alpha\rangle$,
where $|\alpha\rangle$ is the coherent state with the amplitude
$\alpha$ and $a^{\dagger}$ ($a$) is the creation operator
(annihilation operator) of the optical mode.
$N(\alpha,m)=m!L_m(-|\alpha|^2)$, where $L_m(x)$ is the $m$th-order
Laguerre polynomial. When the PACS evolves in the photon loss
channel, the evolution of the density matrix can be described by \cite{Gardiner}\be
\frac{d\rho}{dt}=\frac{\gamma_1}{2}(2a\rho{a}^{\dagger}-a^{\dagger}a\rho-\rho{a}^{\dagger}a),
\ee where $\gamma_1$ represents the dissipative coefficient. The
corresponding non-unitary time evolution density matrix can be
obtained as \beqa
\rho(t)&=&\frac{1}{m!L_m(-|\alpha|^2)}\sum^{\infty}_{k=0}\frac{(1-e^{-\gamma_1{t}})^k}{k!}\nonumber\\
&&\hat{L}(t)a^ka^{\dagger{m}}|\alpha\rangle\langle\alpha|a^ma^{\dagger{k}}\hat{L}(t),\eeqa
where $\hat{L}(t)=e^{-\frac{1}{2}\gamma_1{t}a^{\dagger}a}$.

For the dissipative photon-added coherent state in Eq.(3), the total
output state passing through a 50/50 beam splitter characterized by
the unitary operation
$e^{\frac{\pi}{4}i(a^{\dagger}b+ab^{\dagger})}$ with a vacuum mode b
can be obtained \beqa
\rho_{tot}=D_a(t)D_b(t)e^{-m\gamma_1{t}}e^{|\alpha|^2(e^{-\gamma_1{t}}-1)}\sum^{\infty}_{k=0}\frac{(e^{\gamma_1{t}}-1)^k}{k!}\nonumber\\
\hat{E}^k\hat{E}^{\dagger{m}}|00\rangle\langle00|\hat{E}^m\hat{E}^{\dagger{k}}D^{\dagger}_a(t)D^{\dagger}_b(t),\eeqa
where
$D_a(t)=e^{\frac{\sqrt{2}}{2}\alpha(t){a}^{\dagger}-\frac{\sqrt{2}}{2}\alpha^{\ast}(t){a}}$
and
$D_b(t)=e^{\frac{\sqrt{2}i}{2}\alpha(t){b}^{\dagger}+\frac{\sqrt{2}i}{2}\alpha^{\ast}(t){b}}$
are the displacement operators of the modes a and b, respectively,
where $\alpha(t)=\alpha{e}^{-\gamma_1{t}/2}$.
$\hat{E}=\frac{\sqrt{2}}{2}a-\frac{\sqrt{2}i}{2}b+\alpha{e}^{-\frac{1}{2}\gamma_1{t}}$.
The local unitary operators can not change entanglement, therefore,
we only need to consider the entanglement of the mixed state given
as follows: \beqa
\rho^{\prime}_{tot}=e^{-m\gamma_1{t}}e^{|\alpha|^2(e^{-\gamma_1{t}}-1)}\sum^{\infty}_{k=0}\frac{(e^{\gamma_1{t}}-1)^k}{k!}\nonumber\\
\hat{E}^k\hat{E}^{\dagger{m}}|00\rangle\langle00|\hat{E}^m\hat{E}^{\dagger{k}}.\eeqa
It is obvious that $\hat{E}$ and $\hat{E}^{\dagger}$ satisfy the
commutation relation $[\hat{E},\hat{E}^{\dagger}]=1$. By using this
commutation relation, we can simplify $\rho^{\prime}_{tot}$.
$\hat{E}^k\hat{E}^{\dagger{m}}=\hat{E}^{\dagger}\hat{E}^k\hat{E}^{\dagger(m-1)}+k\hat{E}^{k-1}\hat{E}^{\dagger(m-1)}$.
$\hat{E}^k\hat{E}^{\dagger}|00\rangle=(\alpha(t))^k\hat{E}^{\dagger}|00\rangle+k(\alpha(t))^{k-1}|00\rangle$.
Here, we confine our attention in the case of single quantum
excitation of the classical coherent field, i.e. the single
photon-added coherent state with $m=1$.

For the dissipative SPACS in Eq.(3), the non-classicality of the
evolving density matrix is equivalent to the non-classicality of
superposition of the single-photon Fock state and vacuum state if
the non-classicality is measured by entanglement potential. The
evolving density matrix in Eq.(5) can also be written as \beqa
\rho^{\prime}_{tot}&=&\frac{e^{-\gamma_1{t}}}{1+|\alpha|^2}[\frac{1}{2}|10\rangle\langle10|+\frac{1}{2}|01\rangle\langle01|\nonumber\\
&&+\frac{i}{2}|01\rangle\langle10|-\frac{i}{2}|10\rangle\langle01|+\frac{\sqrt{2}}{2}\alpha^{\ast}e^{\frac{\gamma_1{t}}{2}}|00\rangle\langle10|\nonumber\\
&&-\frac{\sqrt{2}i}{2}\alpha^{\ast}e^{\frac{\gamma_1{t}}{2}}|00\rangle\langle01|+\frac{\sqrt{2}}{2}\alpha{e}^{\frac{\gamma_1{t}}{2}}|10\rangle\langle00|\nonumber\\
&&+\frac{\sqrt{2}i}{2}\alpha{e}^{\frac{\gamma_1{t}}{2}}|01\rangle\langle00|
+(e^{\gamma_1{t}}-1+|\alpha|^2e^{\gamma_1{t}})|00\rangle\langle00|].\nonumber\\
\eeqa The log-negativity of the above density matrix can be
analytically solved as \be
E_p=\log_2(1-\frac{2\chi{e}^{-\gamma_1{t}}}{1+|\alpha|^2}),\ee where
$\chi$ is the unique negative root of the equation \be
8x^3+(4-8e^{\gamma_1{t}}(1+|\alpha|^2))x^2-(6+4e^{\gamma_1{t}}(|\alpha|^2-1))x+1=0.
\ee Though the unique negative root of the above equation can be
analytically obtained, its expression is complicated. From
Eqs.(7-8), it is easy to see that the dissipative SPACS is always
nonclassical for any large but finite decay time. In Fig.1, we have
plotted the log-negativity as the function of $\gamma_1{t}$ and
$|\alpha|$. It is shown that the entanglement potential decreases
with $\gamma_1{t}$, and also decreases with $|\alpha|$ at short
time. In Fig.2, we can see that the entanglement potential exhibits
an exponential decay in the most of the dissipative time. For
$|\alpha|\gg1$, the entanglement potential exhibits an exponential
decay with the loss index \be
E_p(t)\simeq\log_2(1+\frac{e^{-\gamma_1{t}}}{|\alpha|^2+1})\approx\frac{1}{\ln2}\frac{e^{-\gamma_1{t}}}{|\alpha|^2+1}
\ee for any time scale. For $|\alpha|=1$, in short time, \beqa
E_p(t)&=&\log_2[\frac{1}{3}+\frac{1}{6}e^{-\gamma_1{t}}\nonumber\\
&&-\frac{1}{3}\cos(\theta+\frac{2\pi}{3})\sqrt{9e^{-2\gamma_1{t}}+(4-e^{-\gamma_1{t}})^2}],\eeqa
where \be
\theta=\frac{1}{3}\arctan(\frac{3\sqrt{6+3e^{\gamma_1{t}}(12+e^{\gamma_1{t}}(16e^{\gamma_1{t}}-3))}}{-14+e^{\gamma_1{t}}(33+8e^{\gamma_1{t}}(4e^{\gamma_1{t}}-3))}).
\ee However, for long time scale,
$E_p(t)\simeq\log_2(1+\frac{1}{4}e^{-\frac{3}{2}\gamma_1{t}})\simeq\frac{1}{4\ln2}e^{-\frac{3}{2}\gamma_1{t}}$.
For $|\alpha|=0$, the EP of the single-photon Fock state follows a
general curve \be
E_p(t)=\log_2(e^{-\gamma_1{t}}+\sqrt{1-2e^{-\gamma_1{t}}+2e^{-2\gamma_1{t}}})
\ee. Initially, EP decreases linearly with time \be
\gamma_1{t}\ll1~~~E_p(t)\approx1-\frac{\gamma_1{t}}{\ln2},\ee SPACS exhibits the quadrature squeezing when
$|\alpha|>1$. Here, we can obtain the corresponding
relation between the quadrature squeezing and the entanglement
potential of long time dissipative SPACS with $|\alpha|>1$ as \be
E_p(t)\simeq\frac{1-\sigma^2_x}{2\ln2}e^{-\gamma_1t} \ee when
$\gamma_1t\gg\max[1,\ln\frac{2(1+|\alpha|^2)}{(1-|\alpha|^2)^2}]$,
where \beqa
\sigma^2_x&=&\min_{\phi\in[0,2\pi]}\frac{\langle{\Psi_{\alpha,1}}|\hat{x}^2_{\phi}|{\Psi_{\alpha,1}}\rangle-(\langle{\Psi_{\alpha,1}}|\hat{x}_{\phi}|{\Psi_{\alpha,1}}\rangle)^2}{\langle\alpha|\hat{x}^2_{\phi}|\alpha\rangle-(\langle\alpha|\hat{x}_{\phi}|\alpha\rangle)^2}\nonumber\\
&=&\frac{3+|\alpha|^4}{(1+|\alpha|^2)^2} \eeqa is the minimal
variance of quadrature operator
$\hat{x}_{\phi}=\frac{1}{2}(ae^{-i\phi}+a^{\dagger}e^{i\phi})$ of
the initial pure SPACS. However, for the cases with $|\alpha|<1$,
the long time EP is given by \be
E_p(t)\approx\frac{1}{8\ln2}\frac{e^{-2\gamma_1{t}}}{(1+|\alpha|^2)(1-|\alpha|^2)^2},\ee
when
$\gamma_1t\gg\max[1,\ln\frac{2(1+|\alpha|^2)}{(1-|\alpha|^2)^2}]$.
From the above discussions, we understand that the long-time decay
behaviors of the EP of photon-loss SPACSs depend on $|\alpha|$ and a
sharp transition of the decay rate emerges at the critical value $|\alpha|=1$. In
Ref.\cite{Asboth2005}, the EP of the general gaussian squeezing
state in photon loss channel has been derived and shown that the long time
behaviors of EP of general gaussian nonclassical states is
proportional to $e^{-\gamma_1t}$, but not $e^{-2\gamma_1t}$ or
$e^{-\frac{3}{2}\gamma_1t}$. Thus, we can draw the first
remark that photon dissipative SPACSs with initial parameter
$|\alpha|>1$ loss their nonclassicality like the gaussian nonclassical states in the long time scale, while photon dissipative
SPACSs with initial parameter $|\alpha|<1$ loss their nonclassicality like the single-photon Fock state. It will be further conjectured that the
long time decay rates of the nonclassicality are determined by the fact whether their states have quadrature squeezing or not.

We then consider the initial mixed states \be
\rho_M=\frac{p}{1+|\alpha|^2}a^{\dagger}|\alpha\rangle\langle{\alpha}|a+(1-p)|\alpha\rangle\langle\alpha|
\ee in the photon loss channel. These initial mixed states have quadrature
squeezing if $|\alpha|>|\alpha|_c\equiv\frac{1}{\sqrt{2p-1}}$ and $p>\frac{1}{2}$. The non-classicality of $\rho_M$
measured by entanglement potential can be obtained as \be
E_p(\rho_M)=\log_2(1-\frac{2p\tau{e}^{-\gamma_1t}}{1+|\alpha|^2}),
\ee where $\tau$ is the unique negative root of the equation \be
8x^3-(4+8\mu)x^2-(2-4\mu+16\nu)x+1=0,\ee where \beqa
\mu&=&\frac{1}{p}e^{\gamma_1t}(1+|\alpha|^2)-1,\nonumber\\
\nu&=&\frac{1}{2}|\alpha|^2e^{\gamma_1t}.\eeqa It is easy to verify
that the dissipative state is always nonclassical for any large but
finite time. The transition of long time behavior of EP occurs at $|\alpha|=|\alpha_c|=\frac{1}{\sqrt{2p-1}}$ as
$p>\frac{1}{2}$. For $\gamma_1t\gg1$, \beqa
|\alpha|<|\alpha_c|:~~~E_p(t)\propto
e^{-2\gamma_1t}\nonumber\\
|\alpha|=|\alpha_c|:~~~E_p(t)\propto
e^{-\frac{3}{2}\gamma_1t}\nonumber\\
|\alpha|>|\alpha_c|:~~~E_p(t)\propto e^{-\gamma_1t}. \eeqa

Furthermore, we can also derive the \be
\sigma^2_x(p)=1+\frac{2p(1+(1-2p)|\alpha|^2)}{(1+|\alpha|^2)^2} \ee
for the mixed state in Eq.(17), which is smaller than 1 when
$|\alpha|>\frac{1}{\sqrt{2p-1}}$ and $p>\frac{1}{2}$. In this
region, the long time dynamical behavior of EP can be written as \be
\gamma_1t\gg1:~~~E_p(t)\approx\frac{1-\sigma^2_x(p)}{2\ln2}e^{-\gamma_1t},\ee
However, those states in Eq.(17) with
$|\alpha|\leq\frac{1}{\sqrt{2p-1}}$ or $p\leq\frac{1}{2}$ are always out of the neighbor of the
set of gaussian nonclassical states in photon-loss channel. In Fig.3, the $\ln{E_p}$ calculated based on Eqs.(18-20) is depicted as the function of $p$ and $|\alpha|$
at a long decay time $\gamma_1t=9$. It explicitly outlines the transition behavior of the long time decay rate of log-negativity for those states in Eq.(17) with or without quadrature squeezing. It is also consistent with the results in Eq.(21).

In what follows, we pay our attention to the long time decay
behavior of entanglement potential of the cat states in photon loss
channel, which exhibits the squeezing-dependent long time decay rate
too. The cat states are described by \be
|\Psi\rangle=\frac{1}{\sqrt{N}}(|\alpha\rangle+e^{i\phi}|-\alpha\rangle),
\ee where $N$ is the normalization constant
$N=2+2e^{-2|\alpha|^2}\cos(\phi)$. $|\Psi\rangle$ has quadrature
squeezing as $\cos\phi>\cos\phi_c\equiv-e^{-2|\alpha|^2}$. The
dynamical behaviors of EP of the optical field initially prepared in
$|\Psi\rangle$ in the photon loss channel are calculated, and the
results are shown in the Figs.(4-5). In Fig.4, we plot the EP of the
photon loss cat states as the function of $\gamma_1t$ for several
different values of $\phi$ and $|\alpha|$. When
$\cos\phi>-e^{-2|\alpha|^2}$, the initial cat states in Eq.(24)
exhibit the quadrature squeezing and their long time EP can be
derived as \be \gamma_1t\gg1:~~~
E_p(t)\approx\frac{1-\sigma^2_y}{2\ln2}e^{-\gamma_1t},\ee where
$\sigma^2_y=1-\frac{4|\alpha|^2(1+e^{2|\alpha|^2}\cos\phi)}{(e^{2|\alpha|^2}+\cos\phi)^2}$
is the minimal quadrature fluctuation normalized by the ordinary
coherent state. When $\cos\phi<-e^{-2|\alpha|^2}$, the long time EP
is proportional to the $e^{-2\gamma_1t}$. For large $|\alpha|$, at
intermediate times $|2\alpha|^{-2}<\gamma_1t\ll1$, the EP of the cat
states are given by \cite{Asboth2005} \be
E_p(t)\approx\frac{e^{-2\gamma_1t|\alpha|^2}}{(1+e^{-2|\alpha|^2}\cos(\phi))\ln2}.
\ee It can be observed in Fig.4 that the EPs of
the cat states with the same amplitude $|\alpha|$ but different
relative phase $\phi$ firstly decrease with the rule in Eq.(26),
then at a threshold time, the bifurcation occurs and the relative
phase $\phi$ dominates the long time decay patterns of the EPs. In
the inset of Fig.4, the EPs of the cat states at decay time
$\gamma_1t=15$ is plotted as the function of relative phase $\phi$.
The long time decay rate of the EP slightly decreases with $\phi$
from zero to $\frac{\pi}{2}$ and abruptly declines at
$\phi=\arccos(-e^{-2})$. In Fig.5, the natural logarithm of EP at
time $\gamma_1t=10$ is plotted as the function of $|\alpha|$ and
$\phi$. It is shown that the long time decay rate of the EP
undergoes a transition at $\cos\phi_c=-e^{-2|\alpha|^2}$. The
threshold line at which the long time decay rate of the EP undergoes
a transition is just the boundary between quadrature-squeezing
initial cat states and no-quadrature-squeezing initial cat states.

\begin{figure}
\centerline{\includegraphics[width=2.5in]{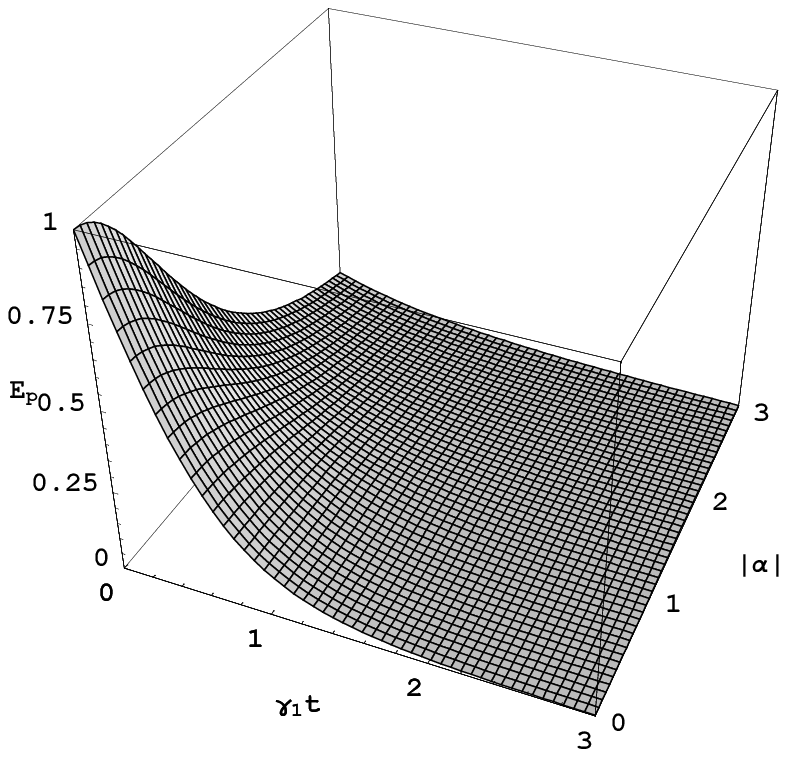}}
\caption{The entanglement potential of the dissipative single
photon-added coherent state is plotted as the function of the
dissipative time $\gamma_1{t}$ and the parameter $|\alpha|$.
Initially, EP decreases linear with time
$E_p(t)=\log_2(\frac{2+|\alpha|^2}{1+|\alpha|^2})-\frac{1}{\ln2}\frac{4+|\alpha|^2}{(2+|\alpha|^2)^2}\gamma_1{t}$.}
\end{figure}

\begin{figure}
\centerline{\includegraphics[width=2.5in]{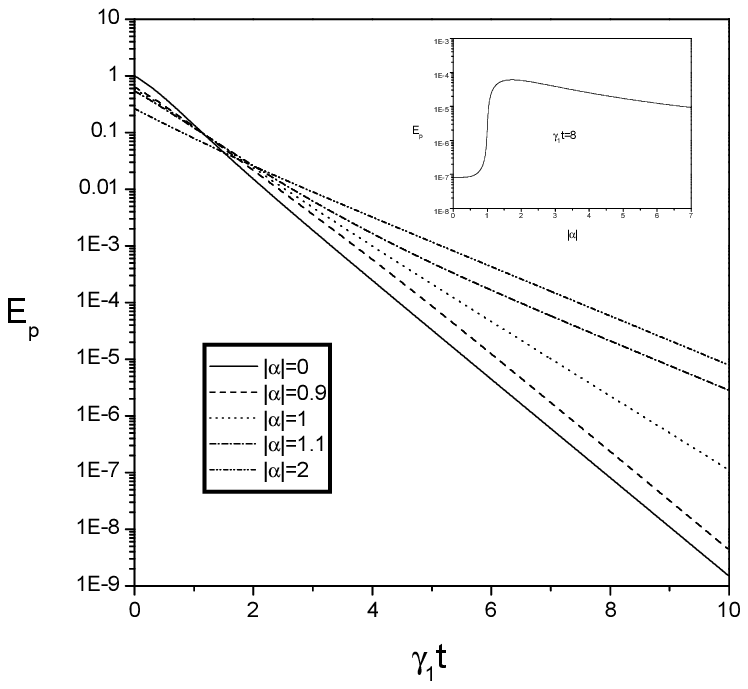}}
\caption{The entanglement potential of the dissipative single
photon-added coherent state is plotted as the function of the
dissipative time $\gamma_1{t}$ for five different values of
$|\alpha|$. It is shown that the non-classicality approximately
exponentially decays with $\gamma_1{t}$ and the decay rate decreases
with $|\alpha|$. From the inset, explicit transition-like behavior
of the long time decay rate of the $E_p$ can be found at
$|\alpha|=1$.}
\end{figure}
\begin{figure}
\centerline{\includegraphics[width=2.5in]{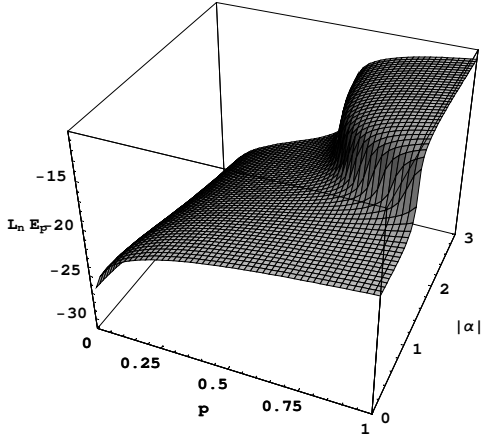}}
\caption{The $\ln{E_p}$ of the mixed states of Eq.(17) in photon loss channel is plotted
as the function of the parameter $p$ and the parameter $|\alpha|$.
$\gamma_1t=9$. }
\end{figure}
\begin{figure}
\centerline{\includegraphics[width=2.5in]{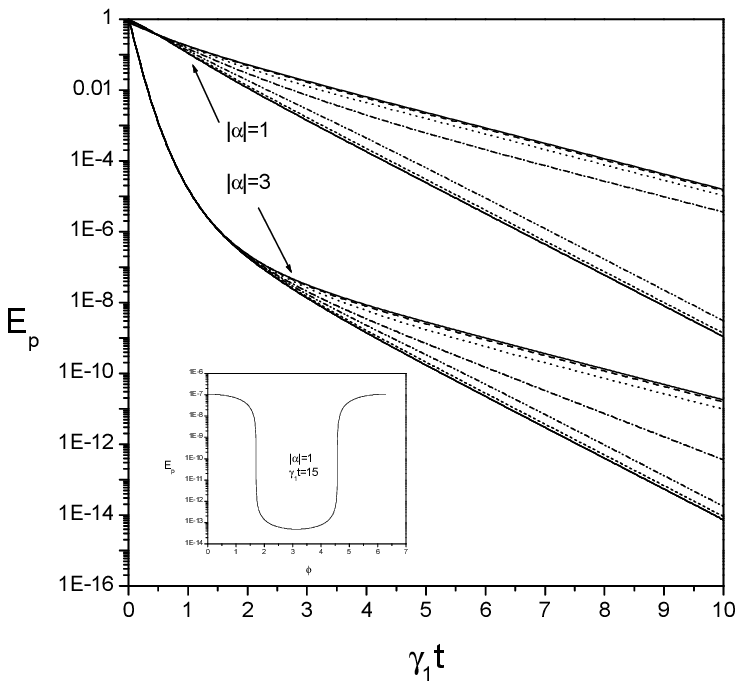}}
\caption{The $E_p$ of the cat states in photon loss channel are
plotted as the function of $\gamma_1t$ with $|\alpha|=1$ for several
different values of the relative phase $\phi$ of the initial states.
For the case with $|\alpha|=1$, (Solid line) $\phi=0$; (Dash line)
$\phi=0.5$; (Dot line) $\phi=1.0$; (Dash dot line) $\phi=1.5$; (Dash
dot dot line) $\phi=2.0$; (Short dash line) $\phi=2.5$; (Short dot
line) $\phi=3.0$. For the case with $|\alpha|=3$, (Solid line)
$\phi=0$; (Dash line) $\phi=0.5$; (Dot line) $\phi=1.0$; (Dash dot
line) $\phi=\pi/2$; (Dash dot dot line) $\phi=2.0$; (Short dash
line) $\phi=2.5$; (Short dot line) $\phi=\pi$. In the inset, the
entanglement potential is shown as the function of the relative
$\phi$ with $\gamma_1t=15$ and $|\alpha|=1$. An abrupt transition
can be found at $\phi\approx1.7$.}
\end{figure}
\begin{figure}
\centerline{\includegraphics[width=2.5in]{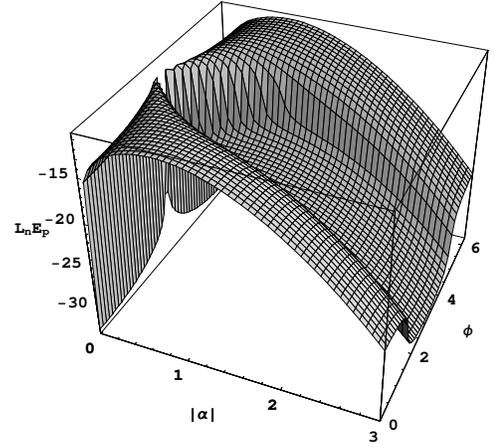}}
\caption{The $\ln(E_p)$ of the cat states in photon loss channel are
plotted as the function of $|\alpha|$ and $\phi$ with
$\gamma_1t=10$. An abrupt decline of the natural logarithm of
entanglement potential at such a long time can be observed at
$\cos\phi=-e^{-2|\alpha|^2}$.}
\end{figure}

\section * {III. NONCLASSICALITY IN THE DEPHASING CHANNEL}

When the PACS evolves in the dephasing channel, the evolution of the
density matrix can be described by \be
\frac{d\rho}{dt}=\frac{\gamma_2}{2}(2{a}^{\dagger}a\rho{a}^{\dagger}a-(a^{\dagger}a)^2\rho-\rho({a}^{\dagger}a)^2),
\ee where $\gamma_2$ represents the dephasing coefficient. The
corresponding non-unitary time evolution density matrix can be
obtained as \beqa
\rho(t)&=&\frac{e^{-|\alpha|^2}}{m!L_m(-|\alpha|^2)}\sum^{\infty}_{k=0}\sum^{\infty}_{n=0}\frac{\alpha^k\alpha^{\ast{n}}\sqrt{(k+m)!(n+m)!}}{k!n!}\nonumber\\
&&e^{-\frac{1}{2}\gamma_2{t}(k-n)^2}|k+m\rangle\langle{n+m}|.\eeqa

For the dephasing photon-added coherent state in Eq.(28), the total
output state passing through a 50/50 beam splitter characterized by
the unitary operation
$e^{\frac{\pi}{4}i(a^{\dagger}b+ab^{\dagger})}$ with a vacuum mode b
can be obtained \beqa
\rho_{tot}=\frac{e^{-|\alpha|^2}}{m!L_{m}(-|\alpha|^2)}\sum^{\infty}_{k=0}\sum^{\infty}_{n=0}\sum^{k+m}_{j=0}\sum^{n+m}_{l=0}\nonumber\\
\frac{(k+m)!(n+m)!\alpha^k\alpha^{\ast{n}}i^{j-l}e^{-\frac{\gamma_2{t}}{2}(k-n)^2}}{k!n!\sqrt{2^{k+n+2m}(k+m-j)!(n+m-l)!j!l!}}\nonumber\\
|k+m-j,j\rangle\langle{n+m-l},l|\eeqa Here, we confine our attention
in the case of single quantum excitation of the classical coherent
field, i.e. the single photon-added coherent state with $m=1$. For
the states in Eq.(29) with $m=1$, the EP is calculated and the
results are displayed in Fig.6. It is shown that, in the dephasing
channel, the non-classicality quantified by EP decreases with the
dephasing time and stays in a stationary value depending on the seed
beam intensity $|\alpha|^2$. The decay rate of the non-classicality
at short time increases with $|\alpha|^2$ which can be explicitly
seen from Fig.7. In the inset, we also plot the total absolute
decrement of EP of SPACS during the dephasing channel, which is
defined by $\Delta{E}_p=E_p(0)-E_p(\infty)$. It can be found that
$\Delta{E}_p$ firstly increases from zero to a maximal value, then
decreases with the further increase of $|\alpha|$. The maximal value
of $\Delta{E}_p$ is achieved at $|\alpha|\approx0.65$. Though SPACS
is considered to be very near the coherent state when $|\alpha|$
tends to very large, the zero-time derivative of non-classicality of
the dephasing SPACS exhibits an incompatible behavior with the
picture that $a^{\dagger}|\alpha\rangle$ tends to coherent state
when $|\alpha|\rightarrow\infty$.

\begin{figure}
\centerline{\includegraphics[width=2.5in]{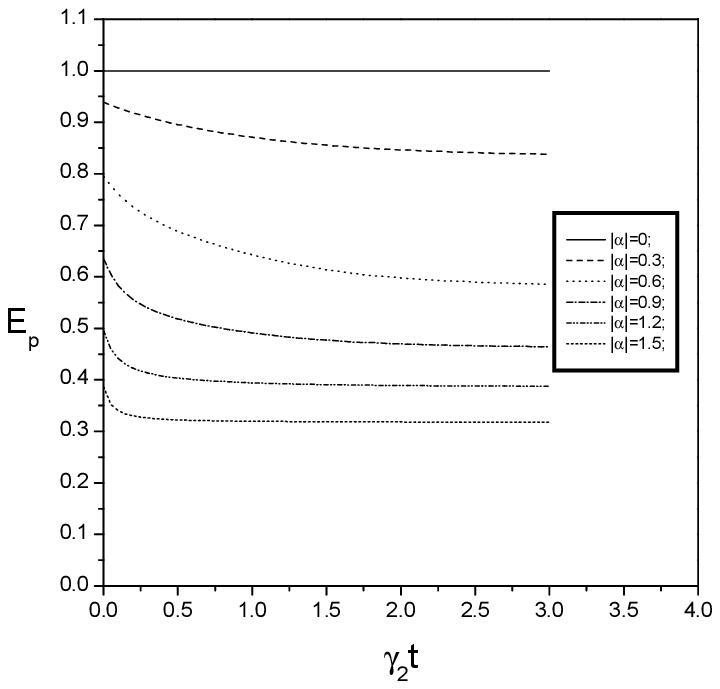}}
\caption{The entanglement potential of the dephasing single
photon-added coherent state is plotted as the function of the
dissipative time $\gamma_2{t}$ for six different values of
$|\alpha|$.}
\end{figure}
\begin{figure}
\centerline{\includegraphics[width=2.5in]{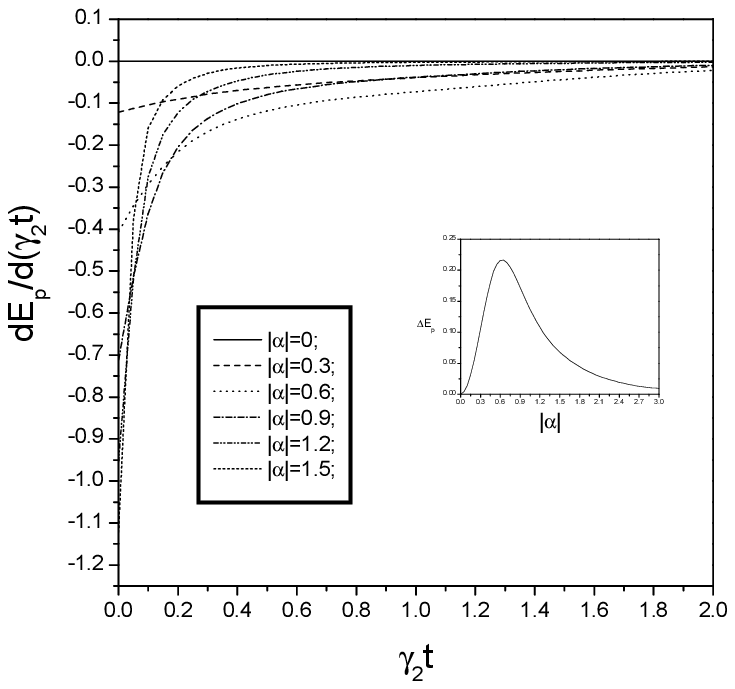}}
\caption{The entanglement potential of the dephasing photon-added
coherent state in Eq.(28) is plotted as the function of the
dephasing time $\gamma_2{t}$ for six different values of $|\alpha|$.
The absolute decrements of EP as the function of $|\alpha|$ are
shown in the inset for the dephasing SPACS.}
\end{figure}

\section * {IV. CONCLUSIONS}

In summary, we have investigated the nonclassicality of several
kinds of nonclassical optical fields such as the pure or mixed
single photon-added coherent states and the cat states in the
photon-loss or the dephasing channels by exploring the entanglement
potential as the measure. It is shown that the long-time decay rates
of entanglement potentials of these states in photon loss channel
are dependent of their initial quadrature squeezing properties. In
the case of the photon-loss, the transition of the long-time decay
rate emerges at the boundary between the squeezing and non-squeezing
initial non-gaussian states if log-negativity is adopted as the
measure of entanglement potential. These examples explicitly show the long time decay of the log-negativity EP is proportional to $\frac{1-\sigma^2}{2\ln2}e^{-\gamma_1t}$ with $\sigma^2$ the minimal quadrature fluctuation normalized by coherent state of the initial states, if the initial states has quadrature squeezing. This result is a generalization of the results in Ref.\cite{Asboth2005} from gaussian squeezing states to nongaussian squeezing states. These examples also show the long time decay of the log-negativity EP of the initial nonsqueezing state in the photon loss channel obeys the decay rule $e^{-2\gamma_1t}$. For the boundary states between squeezing and nonsqueezing, their long time decay of the log-negativity EP obeys the decay rule $e^{-\frac{3}{2}\gamma_1t}$.

However, this kind of transition-like behavior is measure-dependent. We have also adopt the concurrence as the measure of entanglement potential for the case of the SPACS in the photon loss channel, the transition behavior
disappears. Recent studies have revealed very deep relations between the violation of uncertainty relation and the negative partial-transpose states \cite{Nha}. It is conjectured this squeezing-induced transition behavior of long time decay rate of the log-negativity EP exists for more nonclassical nongaussian states in the photon loss channel if their minimal quadrature fluctuation is robust enough against the photon loss.

For the case of the dephasing, distinct
decay behaviors of the nonclassicality are also revealed. For the
dephasing channel, the non-classicality of SPACS is very robust and
the optical field eventually evolves a stationary nonclassical
state. The smaller the
amplitude of initial SPACS, the larger the nonclassicality of the
corresponding stationary state. The decay rate of the non-classicality at short time
increases with $|\alpha|$.

In the Refs.\cite{Li20081}, it has been demonstrated that quantum
excitation of arbitrary optical fields can always exhibit partial
negative Wigner function which will be destroyed and eventually
completely disappear at the same decay time $\gamma{t}_c=\ln2$ in
the photon-loss channel (vacuum environment). Based on the discussion in Appendix A, one can easily draw the following conclusion: If measured by entanglement potential, the nonclassicality of the quantum excitation of any pure single mode optical fields is completely equivalent with the necessary volume of information (quantified by von-Neumann entropy) provided by the vacuum for completely removing the negative Wigner probability distribution.

\section * {APPENDIX A: THE EQUIVALENCE BETWEEN NONCLASSICALITY OF A PURE NONCLASSICAL STATE AND ITS MIXEDNESS ACHIEVED IN PHOTON-LOSS CHANNEL}

In this appendix, we discuss the
equivalence between two concepts for any pure single-mode optical
fields, nonclassicality quantified by entropic entanglement
potential and photon-loss-induced mixedness quantified by
von-Neumann entropy. Assume $EEP$ is the entropic entanglement
potential of an arbitrary single-mode pure state
$|\psi_a\rangle\equiv\sum^{\infty}_{n=0}d_n|n\rangle$. According to
the definition of entropic entanglement potential in
Ref.\cite{Asboth2005}, the $EEP$ of any
single-mode pure states are given by the von-Neumann entropy of the
reduced density operator which is described by \be
\rho_r={\mathrm{Tr}}_{b}[\hat{U}_{BS}|\psi_a\rangle\langle\psi_a|\otimes|0_b\rangle\langle0_b|\hat{U}^{\dagger}_{BS}],\ee
where $\hat{U}_{BS}=e^{\frac{i\pi}{4}(a^{\dagger}b+ab^{\dagger})}$

For any single mode pure states, the measure of
nonclassicality is exactly equivalent to the mixedness of the
quantum field undergoing the photon loss with a fixed characteristic
time in the vacuum environment. For clarifying it, we only need to
recall the equivalence between the master equation describing the photon loss in the
vacuum environment in the interaction picture and the linear loss of beam splitter model \cite{Gerry}. For self-containing of the discussion, we outline its main derivation in what follows. The analytical solution of the master equation describing the photon loss, \be
\frac{\partial\rho}{\partial{t}}=\frac{\gamma}{2}(2a\rho{a}^{\dagger}-a^{\dagger}a\rho-\rho{a}^{\dagger}a),
\ee
can be written as
\be
\rho(t)=\sum^{\infty}_{n=0}\frac{(1-e^{-\gamma{t}})^n}{n!}e^{-\frac{\gamma{t}}{2}a^{\dagger}a}a^n\rho(0)a^{\dagger{n}}e^{-\frac{\gamma{t}}{2}a^{\dagger}a}.
\ee
While the reduced density operator of mode $a$ in the two-mode states produced by the $50/50$ beam splitter injected by the
single mode quantum field in the pure state $\rho(0)\equiv|\psi_a\rangle\langle\psi_a|$ and an auxiliary vacuum
state, can be calculated as
\beqa &&{\mathrm{Tr}}_b[\hat{U}_{BS}(|\psi_a\rangle\langle\psi_a|\otimes|0_b\rangle\langle0_b|)\hat{U}^{\dagger}_{BS}]\nonumber\\
&=&\sum^{\infty}_{n_b=0}\langle{n_b}|\hat{U}_{BS}(|\psi_a\rangle\langle\psi_a|\otimes|0_b\rangle\langle0_b|)\hat{U}^{\dagger}_{BS}|n_b\rangle\nonumber\\
&=&\sum^{\infty}_{n_b=0}\frac{1}{n_b!}\langle{0_b}|b^{n_b}\hat{U}_{BS}(|\psi_a\rangle\langle\psi_a|
\otimes|0_b\rangle\langle0_b|)\hat{U}^{\dagger}_{BS}b^{\dagger{n_b}}|0_b\rangle\nonumber\\
&=&\sum^{\infty}_{n_b=0}\frac{1}{n_b!}\langle{0_b}|\hat{U}_{BS}\hat{B}^{n_b}(|\psi_a\rangle\langle\psi_a|
\otimes|0_b\rangle\langle0_b|)\hat{B}^{\dagger{n_b}}\hat{U}^{\dagger}_{BS}|0_b\rangle\nonumber\\
&=&\sum^{\infty}_{n_b=0}\frac{1}{2^{n_b}n_b!}\langle{0_b}|\hat{U}_{BS}a^{n_b}(|\psi_a\rangle\langle\psi_a|
\otimes|0_b\rangle\langle0_b|)a^{\dagger{n_b}}\hat{U}^{\dagger}_{BS}|0_b\rangle\nonumber\\
&=&\sum^{\infty}_{n_b=0}\frac{1}{2^{n_b}n_b!}e^{-\frac{\ln2}{2}a^{\dagger}a}a^{n_b}|\psi_a\rangle\langle\psi_a|a^{\dagger{n_b}}e^{-\frac{\ln2}{2}a^{\dagger}a},
\eeqa where $\hat{B}=\frac{\sqrt{2}}{2}b+i\frac{\sqrt{2}}{2}a$. In
the derivation of the last step, we have used the formula
$\langle{0_b}|\hat{U}_{BS}|0_b\rangle=\langle{0_b}|\hat{U}^{\dagger}_{BS}|0_b\rangle=e^{-\frac{\ln2}{2}a^{\dagger}a}$.

We can see that the last term in Eq.(33) exactly equal to the
$\rho(t)$ in Eq.(32) with $t=\frac{\ln2}{\gamma}$. The mixedness characterized by the
Von-Neumman entropy of the decohered state
$\rho(\frac{\ln2}{\gamma})$ exactly equals to the two-mode
entanglement produced by the 50/50 beam splitter injected by the
single mode quantum field in the pure state
$\rho(0)\equiv|\psi_a\rangle\langle\psi_a|$ and an auxiliary vacuum
state.

This equivalence
implies that, the larger the nonclassicality of a single mode
quantum field, the more fragile its purity against decoherence
caused by the photon loss in vacuum environment.

\end{document}